\documentstyle[psfig]{mn}

\newif\ifAMStwofonts

\ifoldfss
  \ifCUPmtlplainloaded \else
    \NewTextAlphabet{textbfit} {cmbxti10} {}
    \NewTextAlphabet{textbfss} {cmssbx10} {}
    \NewMathAlphabet{mathbfit} {cmbxti10} {} 
    \NewMathAlphabet{mathbfss} {cmssbx10} {} 
  \fi
  \ifAMStwofonts
    \ifCUPmtlplainloaded \else
      \NewSymbolFont{upmath} {eurm10}
      \NewSymbolFont{AMSa} {msam10}
      \NewMathSymbol{\upi}     {0}{upmath}{19}
      \NewMathSymbol{\umu}     {0}{upmath}{16}
      \NewMathSymbol{\upartial}{0}{upmath}{40}
      \NewMathSymbol{\leqslant}{3}{AMSa}{36}
      \NewMathSymbol{\geqslant}{3}{AMSa}{3E}

       \let\ge=\geqslant
    \fi
  \fi
\fi 

\ifnfssone
  \newmathalphabet{\mathit}
  \addtoversion{normal}{\mathit}{cmr}{m}{it}
  \addtoversion{bold}{\mathit}{cmr}{bx}{it}
  \newmathalphabet{\mathbfit} 
  \addtoversion{normal}{\mathbfit}{cmr}{bx}{it}
  \addtoversion{bold}{\mathbfit}{cmr}{bx}{it}
  \newmathalphabet{\mathbfss} 
  \addtoversion{normal}{\mathbfss}{cmss}{bx}{n}
  \addtoversion{bold}{\mathbfss}{cmss}{bx}{n}
  \ifAMStwofonts
    \ifCUPmtlplainloaded \else
      \UseAMStwoboldmath
      \makeatletter
      \new@mathgroup\upmath@group
      \define@mathgroup\mv@normal\upmath@group{eur}{m}{n}
      \define@mathgroup\mv@bold\upmath@group{eur}{b}{n}
      \edef\UPM{\hexnumber\upmath@group}
      \new@mathgroup\amsa@group
      \define@mathgroup\mv@normal\amsa@group{msa}{m}{n}
      \define@mathgroup\mv@bold\amsa@group{msa}{m}{n}
      \edef\AMSa{\hexnumber\amsa@group}
      \makeatother
      \mathchardef\upi="0\UPM19
      \mathchardef\umu="0\UPM16
      \mathchardef\upartial="0\UPM40
      \mathchardef\leqslant="3\AMSa36
      \mathchardef\geqslant="3\AMSa3E

       \let\ge=\geqslant
    \fi
  \fi
\fi 

\ifnfsstwo
  \DeclareMathAlphabet{\mathbfit}{OT1}{cmr}{bx}{it}
  \SetMathAlphabet\mathbfit{bold}{OT1}{cmr}{bx}{it}
  \DeclareMathAlphabet{\mathbfss}{OT1}{cmss}{bx}{n}
  \SetMathAlphabet\mathbfss{bold}{OT1}{cmss}{bx}{n}
  \ifAMStwofonts
    \ifCUPmtlplainloaded \else
      \DeclareSymbolFont{UPM}{U}{eur}{m}{n}
      \SetSymbolFont{UPM}{bold}{U}{eur}{b}{n}
      \DeclareSymbolFont{AMSa}{U}{msa}{m}{n}
      \DeclareMathSymbol{\upi}{0}{UPM}{"19}
      \DeclareMathSymbol{\umu}{0}{UPM}{"16}
      \DeclareMathSymbol{\upartial}{0}{UPM}{"40}
      \DeclareMathSymbol{\leqslant}{3}{AMSa}{"36}
      \DeclareMathSymbol{\geqslant}{3}{AMSa}{"3E}

       \let\ge=\geqslant
    \fi
  \fi
\fi 

\ifCUPmtlplainloaded \else
  \ifAMStwofonts \else 
    \def\upi{\pi}
    \def\umu{\mu}
    \def\upartial{\partial}
  \fi
\fi
\def\etal{{\it et al. }}

\begin{document}

\title[HST Imaging of the Globular 
Clusters in the Fornax Cluster: NGC 1399 and NGC 1404]
{HST Imaging of the Globular 
Clusters in the Fornax Cluster: NGC 1399 and NGC 1404}

\author[Duncan A. Forbes, Carl J. Grillmair, Gerard M. Williger, 
R. A. W. Elson and Jean P. Brodie]
{Duncan A. Forbes$^1$, Carl J. Grillmair$^2$, Gerard M. Williger$^3$, 
R. A. W. Elson$^4$ 
\newauthor and Jean P. Brodie$^5$\\
$^1$School of Physics and Astronomy, 
University of Birmingham, Edgbaston, Birmingham B15 2TT \\
(E-mail: forbes@star.sr.bham.ac.uk)\\
$^2$Jet Propulsion Laboratory, 4800 Oak Grove Drive, Pasadena, CA 91109, USA\\
(E-mail: carl@grandpa.jpl.nasa.gov)\\
$^3$MPIA, K\"onigstuhl 17, D-69117 Heidelberg, Germany and 
NOAO, Code 681, Goddard Space Flight Center, Greenbelt, MD 20771, USA\\
(E-mail: williger@tejut.gsfc.nasa.gov)\\
$^4$Institute of Astronomy, Madingley Road, Cambridge CB3 0HA\\
(E-mail: elson@star.ast.cam.ac.uk)\\
$^5$Lick Observatory, University of California, Santa Cruz, CA 95064, USA\\
(E-mail: brodie@ucolick.org)}

\pagerange{\pageref{firstpage}--\pageref{lastpage}}
\def\LaTeX{L\kern-.36em\raise.3ex\hbox{a}\kern-.15em
    T\kern-.1667em\lower.7ex\hbox{E}\kern-.125emX}

\newtheorem{theorem}{Theorem}[section]

\label{firstpage}

\maketitle

\begin{abstract}
The Fornax cluster galaxies NGC 1399 and NGC 1404 are ideal for studying
the effects of a cluster environment on globular cluster systems. Here
we present new optical imaging of these two galaxies from both the {\it
Hubble Space Telescope's} Wide Field and Planetary Camera 2 and the Cerro
Tololo Inter--American Observatory's 1.5m telescope. The combination of
both data sets provides unique insight on the spatial and colour
distribution of globular clusters. 
From B--I colours, we find that both galaxies have a broad 
globular cluster metallicity distribution that is inconsistent with a
single population. Two Gaussians provide a reasonable representation of the
metallicity distribution in each galaxy. The 
metal--rich subpopulation is 
more centrally concentrated than the metal--poor one. 
We show that the radial metallicity
gradient can be explained by the changing relative mix of
the two globular cluster subpopulations.  
We derive globular cluster 
surface density profiles, and find that they are flatter
(i.e. more extended) than the underlying starlight. The total number of
globular clusters
and specific frequency are calculated to be 
N = 5700 $\pm$ 500, S$_N$ = 11.5 $\pm$ 1.0 for NGC 1399 and 
N = 725 $\pm$ 145, S$_N$ =
2.0 $\pm$ 0.5 for NGC 1404. 
Our results are compared to the expectations of
globular cluster formation scenarios. 

\end{abstract}




\section{Introduction}

Globular cluster (GC) systems are in some respects 
(e.g. luminosity function),  
more similar to one another than are their
host galaxies.  This suggests an underlying uniformity in the physical
conditions under which GC systems and galaxies formed. On the other hand,
properties such as the total number of clusters per unit galaxy
luminosity show significant variations, and fundamental questions
concerning the origins of these variations remain unanswered. In
particular, the roles of tidal encounters, mergers, cooling flows, and
initial conditions in the formation of GC systems 
have yet to be understood. 

Two galaxies of particular interest in this regard are NGC 1399 at the
center of the Fornax cluster and a nearby cluster member NGC 1404. 
By studying the GC systems of these two galaxies we hope to better
understand the environmental influences on the 
formation of GCs and their host galaxies. 
Various estimates of the distance modulus to the Fornax cluster have been
made in recent years. Here we adopt a typical value of  
m--M = 31.2, which places the cluster 0.2 mag more distant than Virgo (Jacoby
\etal 1992). With this distance modulus and no Galactic extinction
correction, the optical luminosity of NGC 1399 is 
M$_V$ = --21.74 (Faber \etal 1989). This cD galaxy is 
surrounded by hot X--ray emitting gas which extends out
to at least 38$^{'}$ or 190 kpc 
(Mason \& Rosen 1985; Ikebe \etal 1996). 
The mass-to-light ratio increases with
galactocentric distance
reaching a value of M/L $\sim$ 70 M$_{\odot}$/L$_{\odot}$ at $\sim$
5$^{'}$ (Grillmair \etal 1994a). 
Well within the X--ray envelope, even allowing for projection effects, 
at about 10$^{'}$ (50 kpc) to the 
south--east of NGC 1399 lies 
NGC 1404. It is classified as an E1 galaxy and has an optical luminosity
of M$_V$ = --21.37 (Faber \etal 1989).

The GC systems of NGC 1399 and NGC 1404 have been studied using
ground--based imaging by a number of workers (NGC 1399: Hanes \&
Harris 1986; Giesler \& Forte 1990; Wagner \etal 1991; Bridges \etal
1991; Ostrov \etal 1993; Kissler--Patig \etal 1997a and NGC 1404: Hanes
\& Harris 1986; Richtler \etal 1992). For NGC 1399, Kissler--Patig \etal
(1997a) gives the total number of GCs to be 
N$_{GC}$ = 5940 $\pm$ 570. This gives a
high specific frequency S$_N$ = 11 $\pm$ 4. 
A kinematic study of the GCs around NGC 1399 showed that the outer GCs 
appear to
be dynamically related more to the cluster of galaxies than they are to
NGC 1399 itself (Grillmair \etal 1994a). 
This suggests that accretion of intracluster GCs 
(West \etal 1995), tidal stripping (Forte \etal 1982; Forbes, Brodie \&
Grillmair 1997) or mergers (Ashman \& Zepf 1992) may help to explain the
high specific frequency of NGC 1399. 
For NGC 1404 the number of GCs
is less certain. Hanes \& Harris (1986) from a photographic study 
estimated N$_{GC}$ = 190 $\pm$
80, while more recently Richtler \etal (1992) found N$_{GC}$ = 880 $\pm$
120. The resulting S$_N$ values range from 0.5 $\pm$ 0.3 to 2.5 $\pm$
0.3 respectively. The latter value is close to the average for
ellipticals in the Fornax cluster.

Here we present imaging data from {\it HST's} 
Wide Field and Planetary Camera 2 (WFPC2) and the Cerro 
Tololo Inter--American Observatory's (CTIO) 1.5m telescope of NGC 1399
and NGC 1404. The WFPC2 data provide accurate magnitudes and colours
with virtually no contamination from foreground stars or background
galaxies. These data are complemented by wide field-of-view CCD
imaging from CTIO. 
In this paper we will focus on the colour (metallicity) 
and spatial distribution of GCs.  
The GC luminosity functions in these two galaxies and others in the Fornax
cluster are addressed by Grillmair \etal (1997), and a detailed study
of NGC 1379, using WFPC2 and CTIO data, is given by Elson \etal (1997). 

\section{Observations and Data Reduction}

\subsection{HST Imaging}

Details of our HST imaging program are given by Grillmair \etal
(1997). Here we use four pointings from that programme, namely a central
pointing on each of NGC 1399 and NGC 1404, an outer pointing for NGC
1399 and a background field. The outer pointing (F0338)
is situated in the cD envelope of NGC 1399, on the
opposite side of the galaxy from NGC 1404.  
The background field (F0336) is situated
approximately 1.4 degrees south of NGC 1399 in a blank region of sky
and serves to measure the surface density of background sources.

Exposure times totaled 1800s in F814W and 5200s
in F450W for NGC 1399; 1860s and 5000s for NGC 1404 using the WFPC2
camera. 
Three images were taken at one pointing, the other two were offset by
0.5$^{''}$ which corresponds roughly to integer pixel shifts in both the PC
and WFC chips. The images were then aligned and median--combined using
VISTA software which effectively removed both cosmic rays
and hot pixels. The resulting images were found to be statistically
superior to those reduced using standard STSDAS tasks, and 
enabled us to push our completeness limit to about
$B = 26.5$ without incurring large numbers of spurious detections.

Compact sources in each pointing have been detected and measured 
using DAOPHOT II/ALLSTAR software as 
described in Grillmair \etal (1997). For the PC chips the galaxy was
subtracted off first. 
Magnitudes have been converted
into standard Johnson--Cousins B and I. No Galactic extinction correction has
been applied. A colour--magnitude diagram for all of the detected
sources in each of the four pointings is shown in Fig. 1. 
Completeness tests indicate that the 50\% 
incompleteness occurs at 
B $\sim$ 26.5 for all four pointings, with small variations between
each pointing. (Note for the PC chips it is about
0.5 mag brighter.)   
In order to avoid any colour bias
in our results we have excluded all sources fainter than this limit. We have
also excluded sources brighter than B = 21 as these objects would 
be more luminous than Galactic GCs, and so are 
probably foreground stars. We also applied a colour cut to the
data, so that only sources with 1.2 $<$ B--I $<$ 2.5 are
included. This cut is roughly equivalent to --2.5 $<$ [Fe/H] $<$ +1.0, 
assuming the Galactic colour--metallicity relation of Couture \etal
(1990), and covers the range of expected GC metallicities. 
For NGC 1399, 95\% of all sources lie within this colour
selection and for NGC 1404 it is 90\%. This magnitude and colour selection
are indicated in Fig. 1 by dashed lines. With these selection cuts
applied our final sample consists of 572 objects in NGC 1399 central,
59 in NGC 1399 outer  and 208 in NGC 1404. The background field
has only 14 objects. This background field allows us to fairly
accurately estimate the
proportion of {\it bona fide} GCs in the three galaxy pointings, i.e. 
$\ge$ 
98\% in NGC 1399 central, $\ge$ 76\% in NGC 1399 off--center and $\ge$
93\% in NGC
1404. 

\subsection{Ground--based Imaging}

Broad--band B and I images of NGC 1399 and NGC 1404 were taken with the 
CTIO 1.5m telescope. 
We used a Tek 2048 x 2048 array with a pixel
scale of 0.44$^{''}$/pixel, yielding a field-of-view 15$^{'}$ on a side.
The galaxies were observed in 1995
December with typical seeing conditions of $\sim$ 1.5$^{''}$. 
Data reduction was carried out in the standard
way (i.e. bias and
dark subtraction, flat--fielding and sky subtraction). 
The total exposure times were 9600s in B and 3660s 
in I for NGC 1399; 9100s in B and 3060s in I for NGC 1404. 
After combining, the images were
calibrated using aperture photometry from the catalogs
of Longo \& de Vaucouleurs (1983) and de Vaucouleurs \& Longo (1988). 
This procedure gave an rms accuracy of $\pm$ 0.08 mag for NGC 1399 and
$\pm$ 0.05 mag for NGC 1404. No Galactic extinction correction was 
applied.

Globular clusters were detected automatically using DAOPHOT. The
detection threshold was set at a conservative 5$\sigma$ per pixel 
i.e., five times the background noise. 
Selection cuts in the DAOPHOT `sharpness' and `roundness' parameters 
were made to help remove cosmic rays and extended objects 
from the candidate lists. (These
parameters help to remove spikey objects and very
non--round objects.) 
For each detected object we measured a 3 pixel radius
aperture magnitude and applied an aperture correction based on a
curve-of-growth type analysis for a dozen isolated GCs. The rms from
the aperture correction is $\sim$ 0.05 mag. 
Colour--magnitude diagrams for all of the detected
sources for the two galaxies are shown in Fig. 2.
For the CTIO data we 
decided to adopt the same colour selection criteria as the HST data,
and the same bright magnitude cutoff. For the faint magnitude cutoff,
we use the 50\% incompleteness value. Based on an 
examination of the luminosity function, we
estimate that for NGC 1399 the 50\% value is B $\sim$ 24.3 and for NGC
1404 it is B $\sim$ 24.2. Thus our CTIO selection 
criteria are 1.2 $<$ B--I $<$ 2.5 and 21 $<$ B $<$ 24.2/24.3, as is
indicated in Fig. 2 by dashed lines. 
Even with our selection criteria we expect some 
contamination from foreground stars and background
galaxies in our samples of 1752 objects in NGC 1399 and 858 objects 
in NGC 1404. 
This issue will be discussed below. 
We note that although the CTIO data only sample the brighter GCs, we
do not expect this to bias the metallicity distribution.  
For GCs in the Milky Way and M31 (Huchra, Brodie \& Kent 1991) and in 
NGC 5846 (Forbes, Brodie \& Huchra
1996) there appears to be no 
luminosity--metallicity dependence.

\section{Results and Discussion}

\subsection{Spatial Distribution}

Given the magnitude depth and the availability of a background field, our
HST data are ideal to examine the surface density distribution of GCs.
We first calculate the GC density in 5
annuli around NGC 1399. We make use of the fact that out to 100$^{''}$
radius the WFPC2 gives almost complete 180$^{\circ}$ coverage for one
hemisphere of the galaxy (see Forbes, Franx \& Illingworth 
1995 for details). 
For the NGC 1399 outer and background fields we simply add up
the total number of GCs, i.e. 59 and 14 respectively. 
We have also made a small correction ($\sim$ 15\%) for the expected 
number of undetected GCs at the faint end of the luminosity function as
determined by Grillmair \etal (1997). 
Next, we divide the number of GCs by the appropriate spatial coverage
to give a surface density. Finally, we subtract off the background
density of 2.9 objects per square arcmin giving the 
background--corrected surface
density of GCs. This is shown in Fig. 3. 

Excluding the innermost 2 data points for NGC 1399 we have fit the 
data with a function of the form:

$\rho = \rho_{\circ} r^{\alpha}$

\noindent
This fit is shown by a solid line for which $\rho_{\circ}$ = 126
arcmin$^{-2}$ and
$\alpha$ = --1.2 $\pm$ 0.2. 
It is clear from the 
NGC 1399 outer pointing that there are GCs at a projected radius of 8.2
arcmin from NGC 1399. Extrapolating the profile to 9.7 arcmin (the
projected distance of NGC 1404) suggests that there is about one GC per square
arcmin associated with NGC 1399. This corresponds to half a dozen GCs in the
WFPC2 field-of-view centered on NGC 1404. 
The procedure for
estimating the surface density profile of NGC 1404 is similar to
that of NGC 1399, except that we subtract both the background density and
the expected density of GCs in each annulus 
due to NGC 1399. Again the HST data is fit
using a powerlaw profile (excluding the innermost data point) with 
$\rho_{\circ}$ = 36 arcmin$^{-2}$ and $\alpha$ = --1.3 $\pm$ 0.2. 

For both galaxies the GC surface density rises towards the 
galaxy center but flattens off (in log space) in the very inner
regions giving a definable `core' to the GC system where the profile
changes slope.  
This does not appear to be a selection effect and has been
observed in other large galaxies (e.g. Grillmair \etal 1986, 1994b). 
Forbes
\etal (1996) found that the `core radius' of the GC system is loosely
correlated with the galaxy luminosity. 
We estimate that the GC systems of NGC 1399 and NGC 1404 have core
radii of 40$^{''}$ (3.4 kpc) and 30$^{''}$ (2.5 kpc)
respectively. These values are within the scatter of the 
relation found by Forbes \etal (1996) and are similar to the galaxy
effective radii. 
We have measured a GC surface density slope of 
--1.2 $\pm$ 0.2 for NGC 1399. There have been three previous CCD studies
which measured the GC slope. Wagner \etal (1991) found that interior to 
$\sim$ 120$^{''}$ the GC slope was --1.4 $\pm$ 0.11, and this steepened to
--1.54 $\pm$ 0.15 for larger radii. Kissler--Patig \etal (1997a) derived
--1.55 $\pm$ 0.25 and --1.75 $\pm$ 0.3 from two pointings for radii beyond
$\sim$ 50$^{''}$. Bridges \etal (1991) measured --1.4 $\pm$ 0.2 from their
B band data and --1.5 $\pm$ 0.2 from their V band data. 
Our value is somewhat flatter than those measured by others. One possible
reason for this is that our outer pointing is located in the cD envelope of
NGC 1399 where the GC surface density `flattens out' (e.g Kissler--Patig
\etal 1997a). If we exclude the outer point, our fitted slope steepens to
--1.25. Another possibility is that, even though we have not included the
inner two data points in the fit, we are still fitting into the
core--flattened region. 
The stellar profile (log 
intensity) for NGC 1399 is 
also shown in Fig. 3. We measure a slope of --1.6 $\pm$ 0.1 
for the galaxy light. This compares with the measurements of 
--1.67 $\pm$ 0.12 (Wagner \etal 1991) and --1.75 $\pm$ 0.1
(Kissler--Patig \etal 1997a). 
There are been some debate as to whether the GC profile is flatter than the
galaxy profile in NGC 1399. In particular, 
Bridges \etal (1991) claimed that the GC system had the same slope as
the galaxy. Whereas both Wagner \etal
(1991) and Kissler--Patig \etal (1997a) said it was somewhat flatter. 
Excluding our data, the weighted mean value from previous studies is 
--1.47 $\pm$ 0.09 for the GC slope and --1.72 $\pm$ 0.06 for the galaxy
slope. This indicates that the GC profile is 0.25 $\pm$ 0.11 flatter than
the galaxy starlight. Our data suggest that the difference is even greater,
i.e. 0.4 $\pm$ 0.22. Thus, we conclude that the GC surface density profile
is flatter than the galaxy starlight by about 0.3 in the log at about the  
2 $\sigma$ significance level. 
For NGC 1404, we measure a GC slope of --1.3 $\pm$ 0.2 and a 
galaxy starlight slope of 
--1.9 $\pm$ 0.1. 
Thus the GC systems in both galaxies have a
flatter, more extended distribution that the underlying starlight.

Using the surface density profile we can estimate the total number of
GCs and the specific frequency (S$_N$; Harris \& van den Bergh 1981) 
for NGC 1399 and NGC 1404. 
For NGC
1399 GCs are clearly extended beyond 8 arcmin in our data. We have
decided to use 10 arcmin as the limiting radius, which is the same limit
used by Hanes \& Harris (1986) in their photographic study. 
Integrating the powerlaw profile between log r = --0.2 (38$^{''}$) 
and 1 (10 arcmin), and adding the number of GCs interior to 38$^{''}$ gives
a total of 5700 GCs. 
The uncertainty in
deriving the total number of GCs 
is dominated by the choice of limiting radius. We have 
used $\pm$1 arcminute about the limiting radius as 
our error estimate. Thus 
limiting radii of 9 and 11
arcminutes give N = 5700 $\pm$ 500.
This is in good agreement with the determination of Kissler--Patig \etal
(1997a) of N = 5940 $\pm$ 570. 
For an absolute magnitude of M$_V$ = --21.74, the
specific frequency S$_N$ = 11.5 $\pm$ 1.0.

For NGC 1404 the limiting radius is difficult to estimate. Richtler \etal
(1992) determined a background level at $\sim$ 200$^{''}$. We find that in the
direction of NGC 1399, the density of GCs is clearly dominated by NGC 1399
beyond about 250$^{''}$. We have chosen to integrate out to 240$^{''}$
(4 arcmin) with a reasonable range of 3--5 arcminutes. Thus
integrating the profile from log r = --0.3 (30$^{''}$) to 0.6 (4
arcmin) and adding the number of inner GCs gives N = 725 $\pm$ 145 
GCs associated with NGC 1404. This lies  
between the Hanes \& Harris (1986) value (190 $\pm$ 80) and that
of Richtler \etal (1992), i.e. 880 $\pm$ 120. Compared to these
ground--based studies, our HST data has the advantages of accurate
background subtraction (star, galaxies and NGC 1399 GCs) and nearly 
100\% 
complete magnitude coverage. For an absolute magnitude of M$_V$ = --21.37, 
our results indicate a relatively low specific
frequency for NGC 1404 of S$_N$ = 2.0 $\pm$ 0.5. 
Again using the HST data, we calculate the 
`local' S$_N$ values (i.e. using the number of GCs and the integrated
magnitude at the de Vaucouleurs effective radius) to be   
$\sim$ 1.5 for NGC 1399 and $\sim$ 0.5 for NGC 1404. 

In Figures 4 and 5 we show the azimuthal distribution of GCs, as determined
from the HST data, over the range for which we have uniform radial and
azimuthal coverage. 
For NGC 1399 (Fig. 4) 
slight enhancements in the GC counts can be seen close to the galaxy major 
axes (position angles $\sim$ 100$^{\circ}$ 
and --80$^{\circ}$).
There is also a slight deficit around the minor axis P.A. $\sim$
10$^{\circ}$. Thus the GC system is broadly aligned with the stellar
isophotes.  
The direction
towards NGC 1404 (P.A. $\sim$ 150$^{\circ}$) is not covered by our HST
image. 
For NGC 1404, shown in Fig. 5, there is slight deficit in GC counts along
the galaxy
major axis lies at P.A. $\sim$ --20$^{\circ}$. There are also two
enhancements at position angles $\sim$ --50$^{\circ}$ and 10$^{\circ}$ 
which do not correspond to either galaxy major or minor axes.

\subsection{Colour and Metallicity Distributions}

The GC colour distributions, after colour and magnitude selection, 
for the four HST pointings are shown in
Fig. 6. Both NGC 1399 and NGC 1404 are dominated by red (B--I $\sim$
2.1) GCs but with a significant tail to the blue (e.g. B--I $\sim$
1.6). The GCs in the outer
pointing of NGC 1399 are bluer on average than those in the central pointing.
The median colours 
are B--I = 1.97 for NGC 1399 in the central pointing, 1.70
for the outer NGC 1399 pointing and 1.96 for NGC
1404. 

The CTIO data, after colour and magnitude selection, are shown in Fig. 7.
We do not have a background field for the CTIO data. Although we have
attempted to remove galaxies using the DAOPHOT sharpness and roundness
parameters, our seeing
conditions of $\sim$ 1.5$^{''}$ imply that some galaxies will be included
in our candidate list.  
The issue of background galaxy contamination in GC colour/metallicity
distributions is discussed in some detail by Elson \etal (1997). 
Background galaxies in our magnitude range peak at B--I $\sim$
1.0 and get bluer at fainter magnitudes. 
The galaxy counts are falling fast at our blue cutoff of B--I = 1.2. 
Given the richness of
the NGC 1399 GC system, background galaxies are 
likely to be a small affect. However it 
may be more of a concern for NGC 1404.
We can estimate the effects of
background contamination using the CFRS sample (Lilly \etal 1995)
within the appropriate
B magnitude range and scaled in area to match that of the CTIO data. The
background--corrected distributions are shown by a dashed line in Fig. 7. As
we have already removed some galaxies in our initial detection process, the
true distribution may lie somewhere between the two histograms. In either
case (uncorrected or corrected) the distributions 
reveal a somewhat different 
situation to the HST data, with both galaxy's GC systems 
peaking in the blue (B--I $\sim$ 1.6) with a significant red tail
(around B--I $\sim$ 2.1). The median colours 
are B--I = 1.84 for NGC 1399 and 1.71 for NGC 1404. 
Thus the CTIO colours are on average systematically  $\sim$ 0.1 mag bluer 
than the HST data. This result is not formally significant as the error is
also on the order of 0.1 mag, but differences between the shape of the 
two distributions are clearly seen. 
As the HST data cover only the central regions of each
galaxy, this suggests a radial colour gradient from red dominated objects
at small galactocentric radii to blue ones further out. 
In principle, such a gradient could be caused by blue background galaxies in
the CTIO data. However we give further 
evidence below that this is not the case.

In Fig. 8 we show the GC colour distributions for NGC
1399 and NGC 1404 from the HST data. Here the data for NGC 1399 includes both
the central and outer pointing. The background--corrected histograms are
shown by dashed lines. Some galaxies, such as M87 (Whitmore \etal 1995;
Elson \& Santiago 1996) reveal a clear bimodal GC colour distribution. 
In other cases (i.e. NGC 1374, 1379, 1387, 1427) 
the distribution is essentially consistent with a single
unimodal colour for the GC system (Kissler--Patig \etal 1997a,b). The
situation for NGC 1399 and NGC 1404 is not as clear. 
Nevertheless there
is some evidence for multiple GC populations. Firstly, both galaxies have
distributions that are significantly broader than that expected from the
photometric errors (typically $\sigma$ $\sim$ 0.15), 
i.e. there is a real spread in GC colours. 
Secondly, 
we have tested the unbinned colour data using the KMM statistical
test (Ashman, Bird \& Zepf 1994). This test rejects a single Gaussian fit
to the NGC 1399 and NGC 1404 data with a confidence of over 99\%.
If we represent the data with two Gaussians then KMM determines means of 
B--I = 1.7 and 2.1 for NGC 1399, and B--I = 1.6 and 2.1 for NGC 1404.
Thirdly, as an exercise, we generated
a colour histogram that is the sum of two Gaussians. These Gaussians have
mean colours of B--I = 1.6 and 2.1 with dispersions similar to the
photometric error, and with total numbers 
in the ratio of 1:4. They are shown in the lower
left panel of Fig. 8. In the lower right panel, we show the same two
Gaussians but with alternating Poisson noise included. To the eye, the
two noisy Gaussians are qualitatively similar to the HST data, suggesting
they could consist of two Gaussian--like populations. Fourth, the two peaks
indicated in the HST data are also present in the CTIO data (Fig. 7). 
{\it We conclude that both galaxies do not have a single, uniform 
GC population but
rather show evidence for a multimodal GC colour distribution. }

Two other large data sets exist for GCs in NGC 1399. They are the CTIO 4m
observations by Ostrov \etal (1993) using Washington photometry and the
Las Campanas 2.5m observations by Kissler--Patig \etal (1997a) in V and I.
In order to compare all of the different data sets we have converted each
into metallicity. Broad--band colours can be 
transformed into metallicity using the Galactic
relation of Couture \etal (1990) given the usual caveat that age
effects and abundance anomalies may be present. The Washington photometry
of Ostrov \etal is transformed using the relation derived by Geisler \&
Forte (1990). Washington photometry is the most sensitive to metallicity 
(rms $\sim$ 0.1), then B--I colours (rms $\sim$ 0.15) with 
V--I colours being the least sensitive (rms $\sim$ 0.25). In Fig. 9 we show
the GC metallicity distributions from our HST and CTIO data sets, along
with those of Ostrov \etal and Kissler--Patig \etal None of the data sets
have been background corrected in this figure 
(although for the HST sample the correction is negligible). 
In general, the samples reveal a broad
distribution with several minor enhancements, many of which appear common
to several data sets. 
From their data, Ostrov \etal claimed that NGC 1399 has a
trimodal GC metallicity distribution, with 
peaks at [Fe/H] $\sim$ --1.5, --0.8, --0.2. The metal--rich peak
corresponds to that seen clearly in our HST data. The two metal--poor peaks
may be real or simply a single population with [Fe/H] $\sim$ --1.0. 
The mean B--I colours found by the KMM test on the HST data 
correspond to [Fe/H] $\sim$ --1.1 and --0.1. 
The GC metallicity distribution for NGC 1404 from our HST
data is shown in Fig. 10. As indicated by the KMM test, a bimodal
distribution is a better representation of the data than a single
Gaussian. The two GC populations have means of 
[Fe/H] $\sim$ --1.5 and --0.1. 

We now return to the CTIO data and the issue of background contamination. 
Earlier we argued that contamination, although present, did not change the
basic appearance of the colour distributions. Further evidence that the two
peaks in the CTIO data 
are dominated by {\it bona fide} GCs comes from examining the spatial
distribution of the two subpopulations. First, 
we define the metal--rich and metal--poor GC
subpopulations in each galaxy as being $\pm$ 0.3 dex about the mean
metallicity of each subpopulation. 
We then calculate the raw surface density 
(i.e. not corrected for background contamination or for the missing 
faint end of the luminosity function)  for the two
subpopulations. These surface density profiles are shown in Fig. 11.
A background correction would tend to lower the metal--poor profile,
particularly at galactocentric radii beyond 4 arcmin in NGC 1404 
(which is excluded from the figure).
Although
there is considerable scatter, the surface density of 
each subpopulation declines with distance from its parent galaxy. This
indicates that each subpopulation is associated with the galaxy and is 
therefore dominated by GCs. 
Secondly, the 
fact that the HST colours (metallicities)
agree with those in the inner annuli of the CTIO data (see figures 12
and 13) indicates that the subpopulations are dominated by GCs. 

Next we investigate the 
radial variation of GC metallicity.
We have seen that for 
both NGC 1399 and NGC 1404, the metal--rich (red) subpopulation 
dominates in the HST data and in the CTIO
data the metal--poor (blue) subpopulation is dominant.
As the HST data probe the inner $\sim$
100$^{''}$ region and the CTIO data cover the 
regions beyond $\sim$ 30$^{''}$ 
this would suggest that the metal--rich GCs are more centrally
concentrated than the metal--poor ones. 
A similar situation is inferred for NGC 4472 (Geisler
\etal 1996) and NGC 5846 (Forbes, Brodie \& Huchra 1997).
An unweighted fit to the HST data indeed 
indicates a metallicity gradient in both galaxies. 
For NGC 1399 we find [Fe/H] = --0.31$\pm$0.13 log R (arcsec) --
0.046$\pm$0.22. This slope is similar to that found by Ostrov \etal (1993)
i.e. --0.34$\pm$0.18. The fit  
for NGC 1404 gives [Fe/H] = 
--0.36$\pm$0.13 log R (arcsec) --
0.034$\pm$0.22. In both cases the HST fits are consistent with the
CTIO data. 
In NGC 4472, 
Geisler \etal (1996) found that the radial GC 
metallicity gradient was actually due to the changing relative mix of the
two GC subpopulations. Assuming that NGC 1399 and NGC 1404 can be
represented by metal--rich and metal--poor GC subpopulations we investigate
the radial gradients further. 

In Figures 12 and 13 we show the metallicity in various radial bins for the
different GC subpopulations and the mean metallicity for the combined
population. The range in metallicity included in each data point is $\pm$
0.3 dex.  
For NGC 1399 (Fig. 12) the metal--poor and 
metal--rich GC subpopulations show no obvious radial metallicity
gradient. However the mean values for the whole GC system {\it do} show a
global radial gradient. Although not quite as
convincing, the same trend appears to be true for NGC 1404 (Fig. 13). 
For both galaxies, we conclude that the 
radial metallicity gradient for the overall GC system is consistent 
with the changing relative proportions of the GC 
subpopulations; the metal--rich GCs are centrally concentrated
while the metal--poor ones are preferentially
located in the outer regions.  The lack of a real
abundance gradient places constraints on the role of gas dissipation in the
formation of GCs. 
In NGC 4472, the metallicity of the galaxy field stars was 
found to be of the same or of slightly higher metallicity than the
metal--rich GCs (Geisler \etal 1996). 
For NGC 1399, Brodie \& Huchra (1991)
quote a
spectroscopic metallicity of [Fe/H] = +0.2 $\pm$ 0.9. This is similar
to the metallicity derived from B--I colours of the galaxy (from our 
CTIO data and 
from Goudfrooij \etal 1994). For NGC 1404, the galaxy B--I colours 
indicate [Fe/H]
$\sim$ --0.2. Thus for both galaxies, the field stars appear to have
similar metallicities to those of metal--rich GCs (which have [Fe/H] $\sim$
--0.1). 

\subsection{Comparison with Globular Cluster Formation Scenarios}

\subsubsection{The Merger Model}

The merger model of Ashman \& Zepf (1992) and Zepf \& Ashman (1993)
make some specific predictions concerning the properties of GCs in 
ellipticals that are the result of a gaseous merger. One of their key
predictions is that ellipticals will have multimodal GC metallicity
distributions. Furthermore the metal--rich GCs will be centrally
concentrated and the more metal--poor ones will be preferentially 
located at large galactocentric radii. This indeed appears to be the
case for both NGC 1399 and NGC 1404,  
and is therefore in qualitative agreement with their model. For NGC 1404
with a low S$_N$ value (i.e. 2.0 $\pm$ 0.5), Ashman \& Zepf (1992)
expect slightly fewer metal--rich (new) than metal--poor
(old) GCs. Our data are generally consistent with this expectation.  
For NGC 1399 with S$_N$ = 11.5 $\pm$ 1.0, they would expect N$_R$/N$_P$
$\sim$ 3--4. In the central regions of NGC 1399 such a ratio
may hold, but globally N$_R$/N$_P$ $<$ 1. This is of course complicated by
the difficulty in defining clear metal--rich and metal--poor GC
subpopulations. However there seems to be a 
shortfall in the number of newly created GCs if 
gaseous mergers are responsible for the metal--rich GCs and the high S$_N$
value.  
Another expectation from their merger model is that
the galaxy starlight and GC system should have a similar spatial
distribution in galaxies with 
high S$_N$ values (i.e. those galaxies in which large 
numbers of new GCs should have been created by the merger). 
Thus we might expect the surface density of
GCs to match that of the galaxy surface brightness profile for NGC 1399 but
not for NGC 1404. 
Our data indicate that for both galaxies the GC distribution is notably
flatter than the underlying starlight. 
We might also expect the metal--rich GCs, which should have
formed from the gas of the progenitor galaxies, to reveal a radial
metallicity gradient as the formation process should be dissipative. 
Figures 12 and 13 suggest that there is little or no metallicity gradient
for the metal--rich GCs in either galaxy.

To summarize, the GC systems in NGC 1399 and NGC 1404 reveal some
properties that are consistent with the Ashman \&
Zepf (1992) merger model, but there are also notable conflicts.
On an individual basis, some of the disagreements with the model
may be accommodated by slight modifications to the initial model
(see Zepf, Ashman \& Geisler 1995). 
Forbes, Brodie \& Grillmair 
(1997) have recently investigated
whether a sample of elliptical galaxies meets the general predictions
of the Ashman \& Zepf (1992) model, and  
concluded that mergers were
{\it unlikely} to account for the GC systems in large elliptical galaxies. 
They favored a multiphase collapse scenario. Next we compare the data
for NGC 1399 and NGC 1404 with this scenario. 

\subsubsection{The Multiphase Collapse Model}

Forbes, Brodie \& Grillmair (1997) have proposed that GCs in
ellipticals form {\it in situ} during a multiphase collapse. In the
first pre--galactic phase, metal--poor GCs are formed and later on, 
in the second galactic phase, the metal--rich GCs form. This leads to
a bimodal metallicity distribution with the metal--rich GCs more
centrally concentrated than the metal--poor ones.  In massive ellipticals 
the collapse could be largely
dissipationless and so no strong radial abundance gradients are
expected. The radial distribution of (the metal--poor) 
GCs will be more extended
than the galaxy itself, if as expected, 
they form before the galactic stars. 
The overall metallicity and spatial properties of GCs 
derived from our data are
consistent with this picture. 

For cD galaxies, such as NGC 1399, Forbes, Brodie \& Grillmair (1997) 
suggested an additional source of GCs from
tidal stripping of nearby galaxies. They suggested that some of GC 
subpopulation identified by Ostrov \etal (1993) 
at around [Fe/H] $\sim$ --0.8 were
acquired from NGC 1404 and possibly other Fornax galaxies. 
Recently, Jone \etal (1997) have interpreted the asymmetric X--ray emission
around NGC 1404 as evidence for tidal interaction with NGC 1399. 
The idea of tidal stripping of GCs is
given some support by our new data. 
The outer regions of NGC 1404 are dominated by metal--poor GCs, and 
it is these
GCs that may be preferentially stripped and captured by NGC 1399. Thus some
of the GCs in NGC 1399 with this intermediate metallicity may 
have originated from NGC 1404. 
We confirmed that NGC 1404 has a remarkably low specific frequency 
value (S$_N$ = 2.0 $\pm$ 0.5) for a cluster elliptical, 
suggesting that some GCs may have been lost. 
If NGC 1404 originally had S$_N$ = 5, i.e. typical of normal ellipticals
outside of the Fornax cluster then it
has lost $\sim$ 1000 GCs, which may have been  
tidally captured by NGC 1399. 
If we assume that the GC system and galaxy starlight originally
(i.e. before tidal stripping) continued beyond 4 arcminutes with the same
slope as found in section 3.1, then 
we can calculate the S$_N$ value of the stripped GCs. 
We find that the stripped GCs are unlikely reach values of S$_N$
$\sim$ 11.5 in sufficient numbers to have any appreciable effect on the
specific frequency of NGC 1399.  
We expected
the GC surface density profile to match that of the underlying
starlight in the inner regions of NGC 1404, 
but it appears to be somewhat flatter than the stellar
profile (at least for galactocentric radii of less than $\sim$
150$^{''}$). This may just be indicating that tidal stripping has not yet
affected the galaxy inner regions. The above 
circumstantial evidence suggests that NGC 1399
has acquired GCs from NGC 1404, although this is unlikely to explain the
high specific frequency of NGC 1399. 
Large samples of GC metallicity and 
kinematic information from spectra are probably required before the
multiphase collapse and tidal stripping ideas 
can be fully tested.

\section{Conclusions}

We present new optical data of the globular cluster 
(GC) systems in NGC 1399 and NGC 1404. 
Our images were taken in B and I filters 
using the Hubble Space Telescope (HST) and the 1.5m telescope at 
Cerro Tololo Inter--American Observatory (CTIO). This provides us with high
spatial resolution in the central regions and wide area coverage
respectively. From the HST data we have detected over 500 
GCs in NGC 1399 and over 200 in NGC 1404, covering $\sim$ 90\% of the
GC luminosity function. With the aid of an HST background field of
the same exposure time, we
estimate that the contamination levels in the HST samples are only a few
percent. 

Both galaxies reveal a broad GC colour/metallicity distribution. These
distributions are {\it inconsistent} with a single Gaussian. For NGC 1399, 
two Gaussians with mean metallicities of [Fe/H] $\sim$ --1.1 and --0.1
provide a reasonable representation of the data, although a multimodal
distribution with several different GC subpopulations remains a
possibility. In the case of NGC 1404 we tentatively claim evidence for a
bimodal GC distribution with the 
metal--poor population around [Fe/H] $\sim$ --1.5 and the metal--rich
population at [Fe/H] $\sim$ --0.1. 
In both galaxies
the different metallicity peaks are consistent with distinct GC 
subpopulations in which the metal--rich
GCs are more centrally concentrated than the metal--poor ones.
Globally, the GC systems reveal metallicity gradients. However 
these gradients are consistent with a changing relative mix of the
two GC subpopulations with galactocentric radius. 

We derive GC surface density profiles, and show them to be flatter
(i.e. more extended) than the underlying starlight. The total number of GCs
and specific frequency are calculated to be 
N = 5700 $\pm$ 500, S$_N$ = 11.5 $\pm$ 1.0 for NGC 1399 and 
N = 725 $\pm$ 145, S$_N$ =
2.0 $\pm$ 0.5 for NGC 1404. 
Finally we discuss our results in the context
of two GC formation scenarios. The GC data on NGC 1399 and NGC 1404 
are generally more consistent
with a multiphase collapse (e.g. Forbes, Brodie \& Grillmair 1997) than a
merger (e.g. Ashman \& Zepf 1992) origin.

\noindent
{\bf Acknowledgments}\\
We thank M. Kissler--Patig and M. Rabban for help and 
useful discussions. We also thank the referee, T. Bridges, for several
helpful comments. This
research was funded by the HST grant GO-05990.01-94A.\\

\newpage
\noindent{\bf References}

\noindent
Ashman, K. M., Bird, C. M., Zepf, S. E., 1994, AJ, 108, 2348\\
Ashman, K. M., Zepf, S. E., 1992, ApJ, 384, 50\\
Brodie, J. P., Huchra, J., 1991, ApJ, 379, 157\\
Bridges, T. J., Hanes, D. A., Harris, W. E., 1991, AJ, 101, 469\\
Couture, J., Harris, W. E., Allwright, J. W. B., 1990, ApJS, 73,
671\\
de Vaucouleurs, A., Longo, G., 1988, Catalogue of visual and infrared
photometry of galaxies from 0.5$\mu$m to 10$\mu$m. University of
Texas, Austin\\
Elson, R. A. W., Santiago, B. X., 1996, MNRAS, 280, 971\\
Elson, R. A. W., Grillmair, C. J., Forbes, D. A., Rabban, M., Williger,
G. M., Brodie, J. P., 1997, MNRAS, submitted\\
Faber, S. M., \etal 1989, ApJS, 69, 763\\
Forbes, D. A., Brodie, J. P., Huchra, J., 1996, AJ, 112, 2448\\
Forbes, D. A., Brodie, J. P., Huchra, J., 1997, AJ, 113, 887\\
Forbes, D. A., Brodie, J. P., Grillmair, C. J.., 1997, AJ, 113, 1652\\
Forbes, D. A., Franx, M., Illingworth, G. D., Carollo, C. M., 1996,
ApJ, 467, 126\\
Forbes, D. A., Franx, M., Illingworth, G. D., 1995, AJ, 109, 1988\\
Forte, J. C., Martinez, R. E., Muzzio, J. C., 1982, AJ, 87, 1465\\
Geisler, D., Lee, M. G., Kim, E., 1996, AJ, 111, 1529\\
Geisler, D., Forte, J. C., 1990, ApJ, 350, L5\\
Goudfrooij, P., Hansen, L., Jorgensen, H. E., Norgaard Nielson, H. U.  
de Jong, T., van den Hoek, L. B., 1994, A \& AS, 104, 179\\
Grillmair, C., \etal 1994a, ApJ, 422, L9\\
Grillmair, C., \etal 1994b, AJ, 108, 102\\ 
Grillmair, C., Pritchet, C., van den Bergh, S., 1986, AJ 91, 1328\\
Grillmair, C., Forbes, D. A., Brodie, J. P.,  
Elson, R. A. W., 1997, AJ, submitted\\
Hanes, D. A., Harris, W. E., 1986, ApJ, 309, 564\\
Harris, W. E., van den Bergh, S., 1981, AJ, 86, 1627\\
Huchra, J., Brodie, J. P., Kent, S., 1991, ApJ, 370, 495\\
Ikebe, Y., \etal 1996, Nature, 379, 427\\
Jacoby, G. H., \etal 1992, PASP, 104, 599 (J92)\\
Jones, C., Stern, C., Forman, W., Breen, J., David, L., Tucker, W., \&
Franx, M. 1997, ApJ, 482, 143\\
Kissler--Patig, M., \etal 1997a, A \& A, 319, 470\\ 
Kissler--Patig, M., Forbes, D. A., Minniti, D., 1997b, in preparation\\
Kohle, S., \etal 1995, A \& A, 309, L37\\
Lilly, S. J., Le Fevre, O., Crampton, D., Hammer, F., Tresse, L., 1995, ApJ,
455, 50\\
Longo, G., de Vaucouleurs, A., 1983, A General Catalogue of
Photometric Magnitudes and Colours in the UBV system. University of
Texas, Austin\\
Madore, B., \etal 1997, in preparation\\
Mason, K. O., Rosen, S. R., 1985, Sp. Sc. Rev., 40, 675\\ 
Ostrov, P., Geisler, D., Forte, J. C., 1993, AJ, 105, 1762\\
Richtler, T., Gerbel, E. K., Domgorgen, H., Hilker, M., Kissler, M., 1992,
A \& A, 264, 25\\
Wagner, S., Richtler, T., Hopp, U., 1991, A \& A, 241, 399\\
West, M. J., Cote, P., Jones, C., Forman, W., Marzke, R. O., 1995,
ApJ, 453, L77\\
Whitmore, B. C., Sparks, W. B., Lucas, R. A., Macchetto, F. D., 
Biretta, J. A., 1995, ApJ, 454, L73\\
Zepf, S. E., Ashman, K. M., 1993, MNRAS, 264, 611\\
Zepf, S. E., Ashman, K. M.,  Geisler, D., 1995, ApJ, 443, 570\\


\newpage

\begin{figure*}
\centerline{\psfig{figure=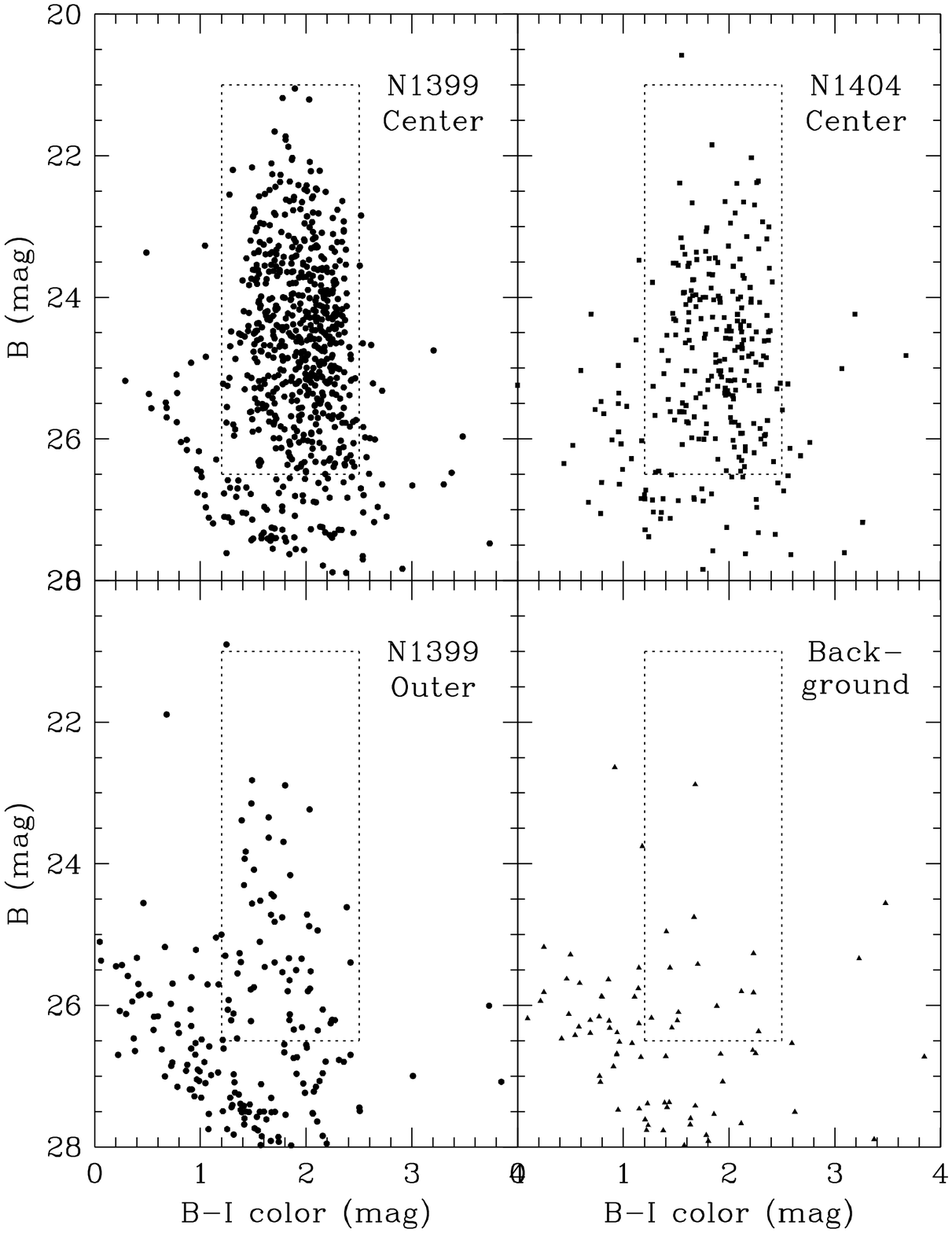,width=400pt}}
\caption{\label{fig1}
Colour--magnitude diagrams for the compact objects in four HST
pointings: central NGC 1399 (circles), central NGC 1404 (squares), 
outer NGC 1399 (circles) and a
background field (triangles). The dashed box indicates the colour and magnitude
selection criteria (see text for details).
}
\end{figure*}

\begin{figure*}
\centerline{\psfig{figure=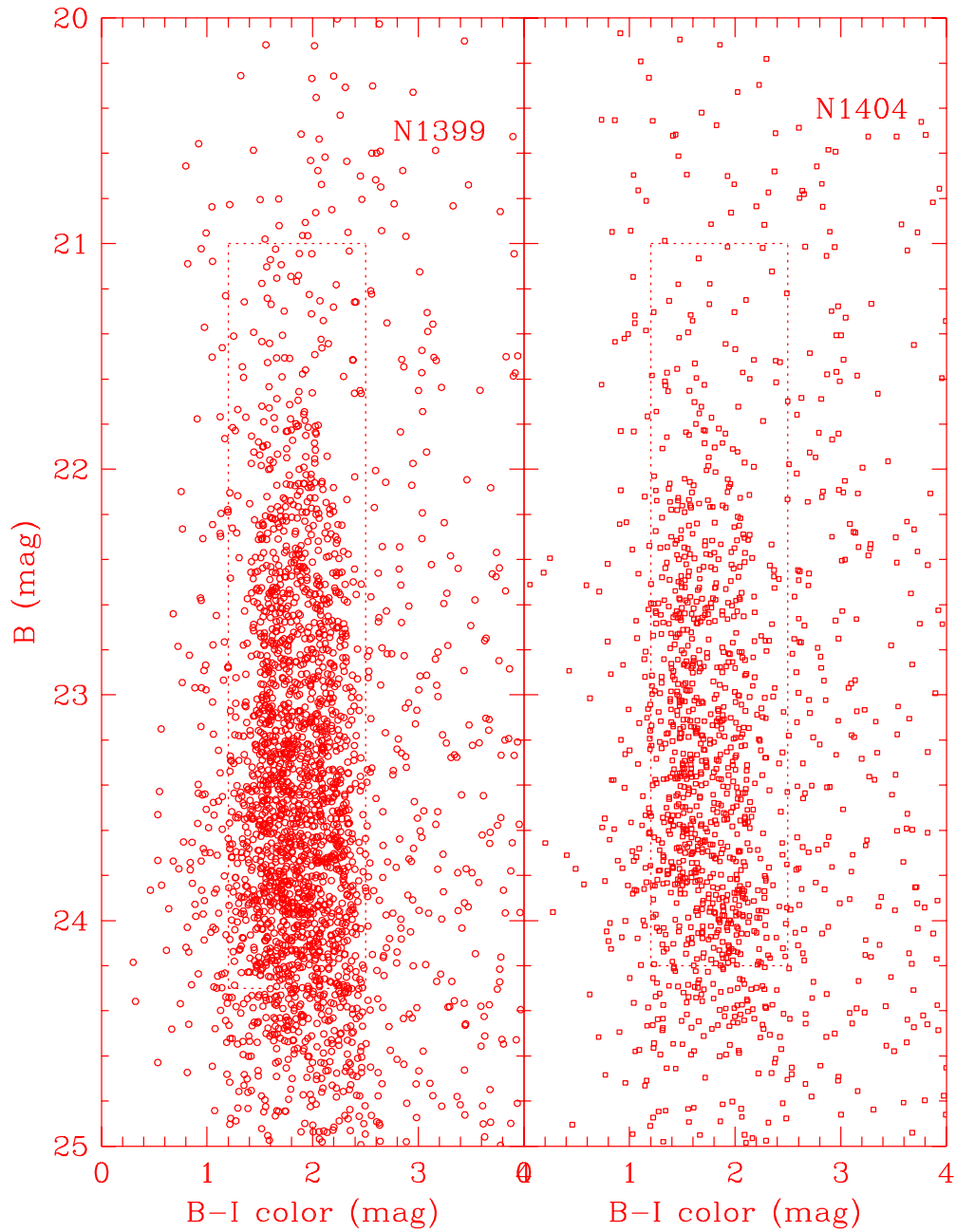,width=400pt}}
\caption{\label{fig2}
Colour--magnitude diagrams for the compact objects in two CTIO 
pointings centered on NGC 1399 (circles) and NGC 1404 (squares).
The dashed box indicates the colour and magnitude
selection criteria (see text for details).
}
\end{figure*}

\begin{figure*} 
\centerline{\psfig{figure=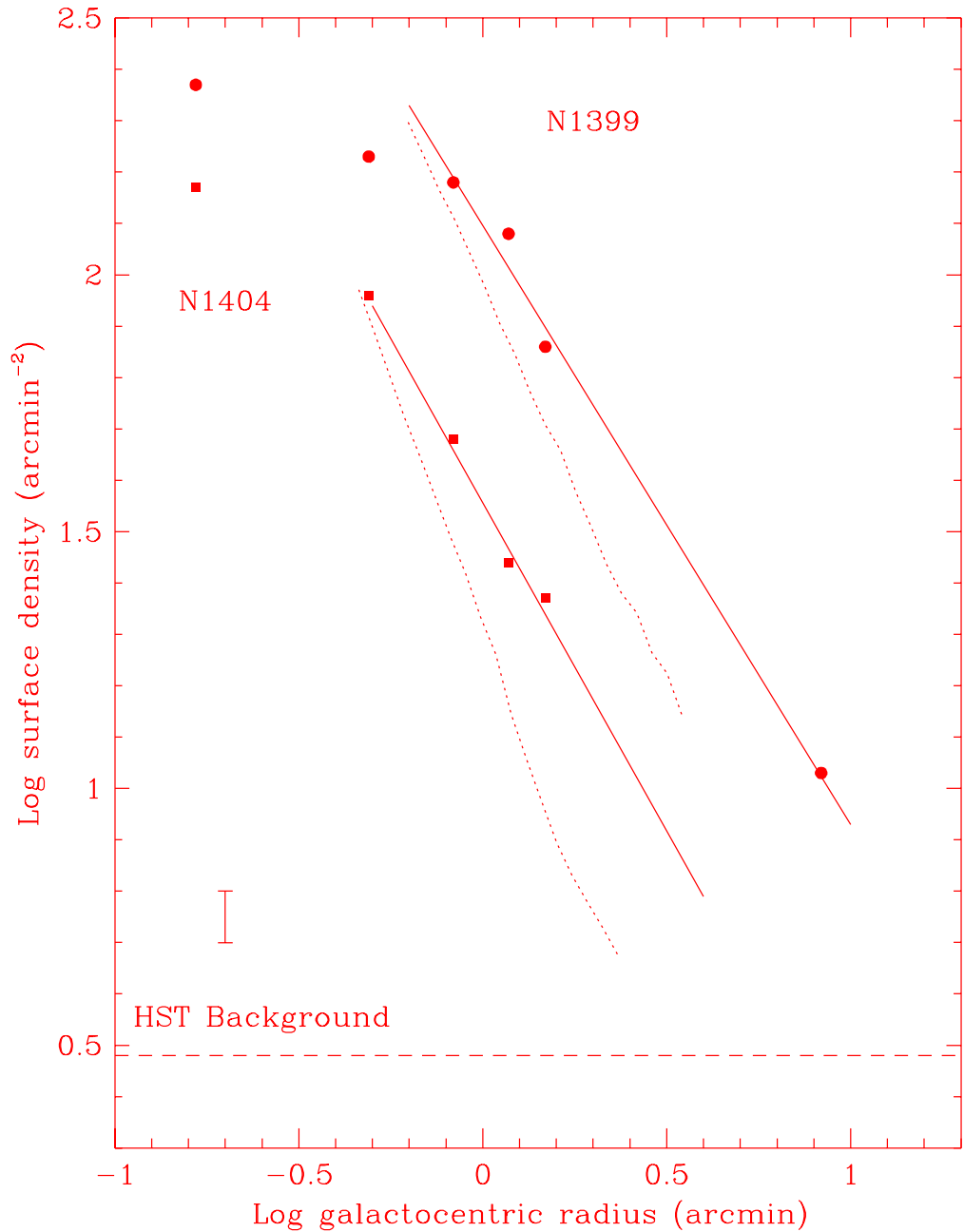,width=400pt}}
\caption{\label{fig3}
Surface density profiles for the globular cluster systems in NGC 1399 
(circles) and NGC 1404 (squares) from 
HST data. The surface densities have had a correction for 
background contamination and 
for the missing faintest globular clusters. 
The globular cluster systems of both galaxies reveal a `core region'. 
Powerlaw fits to the data beyond the core region 
are shown by solid lines. The surface density of compact
background objects in HST images is shown by a long--dashed line.    
A typical error bar is shown on the left.
The underlying galaxy starlight profile, arbitrarily normalized in the
vertical direction, is shown by a short--dashed line. For both galaxies the
starlight is more concentrated than the globular cluster system.
}
\end{figure*}

\begin{figure*} 
\centerline{\psfig{figure=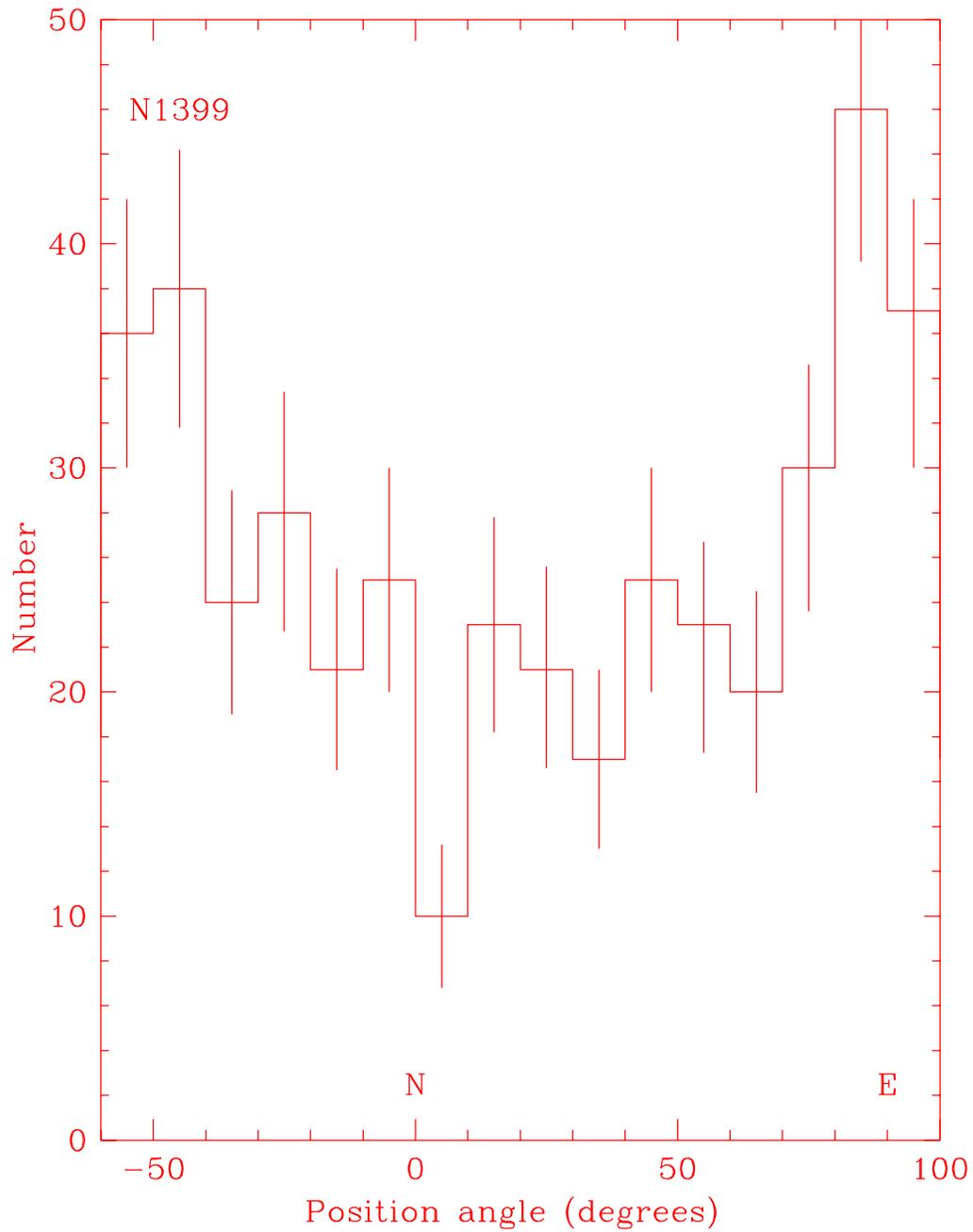,width=400pt}}
\caption{\label{fig4}
Histogram of globular cluster position angles within 100$^{''}$ radius of
NGC 1399, from HST data. 
The galaxy major axes are P.A. $\sim$ 100$^{\circ}$ and --80$^{\circ}$. 
Globular clusters
appear to be concentrated close to the galaxy major axis and show a
deficit along the minor axis. 
}
\end{figure*}

\begin{figure*} 
\centerline{\psfig{figure=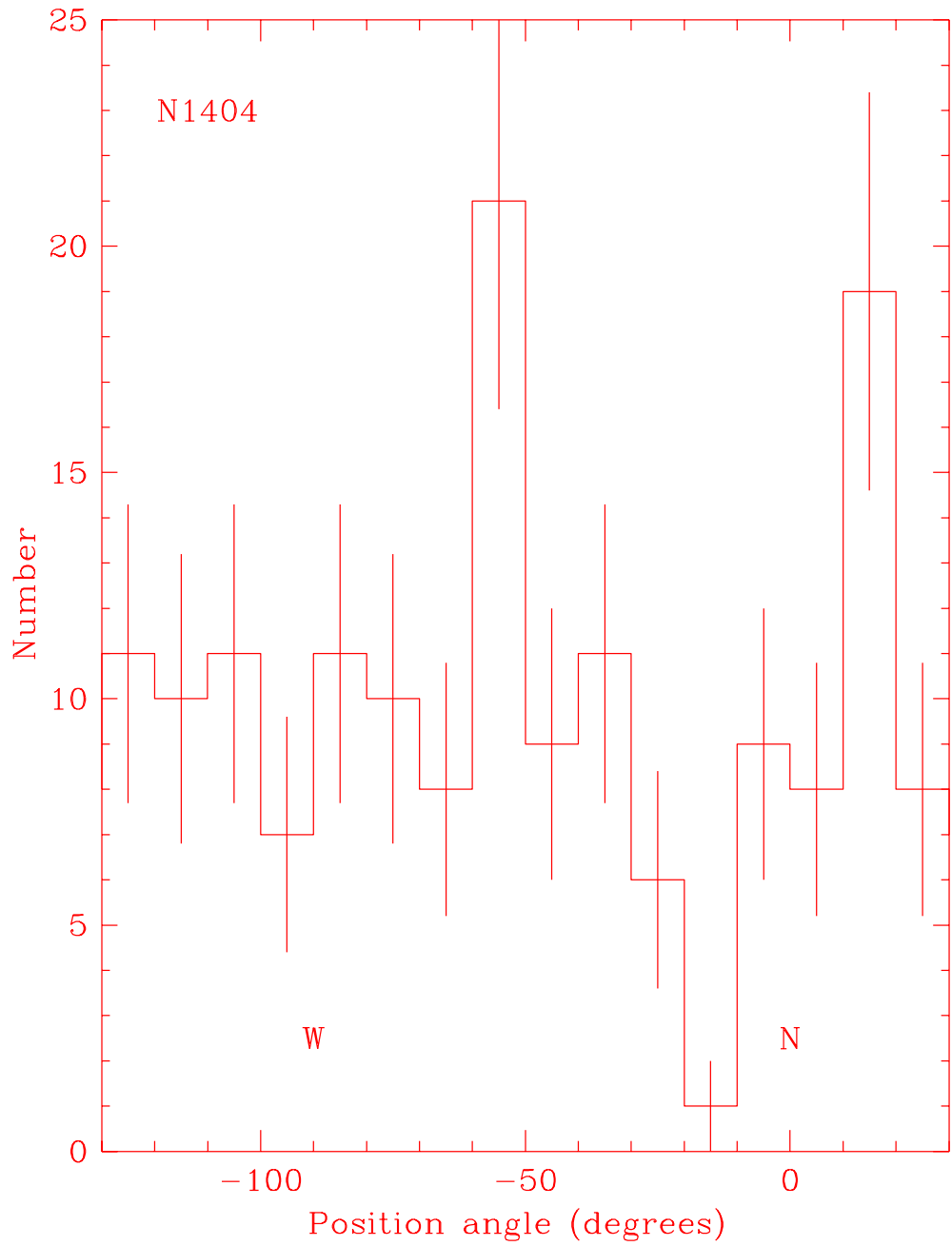,width=400pt}}
\caption{\label{fig5}
Histogram of globular cluster position angle within 100$^{''}$ of NGC
1404, from HST data. 
The galaxy major axis is P.A. $\sim$ --20$^{\circ}$. Globular clusters
appear to show a slight deficit near the galaxy major axis.
}
\end{figure*}

\begin{figure*} 
\centerline{\psfig{figure=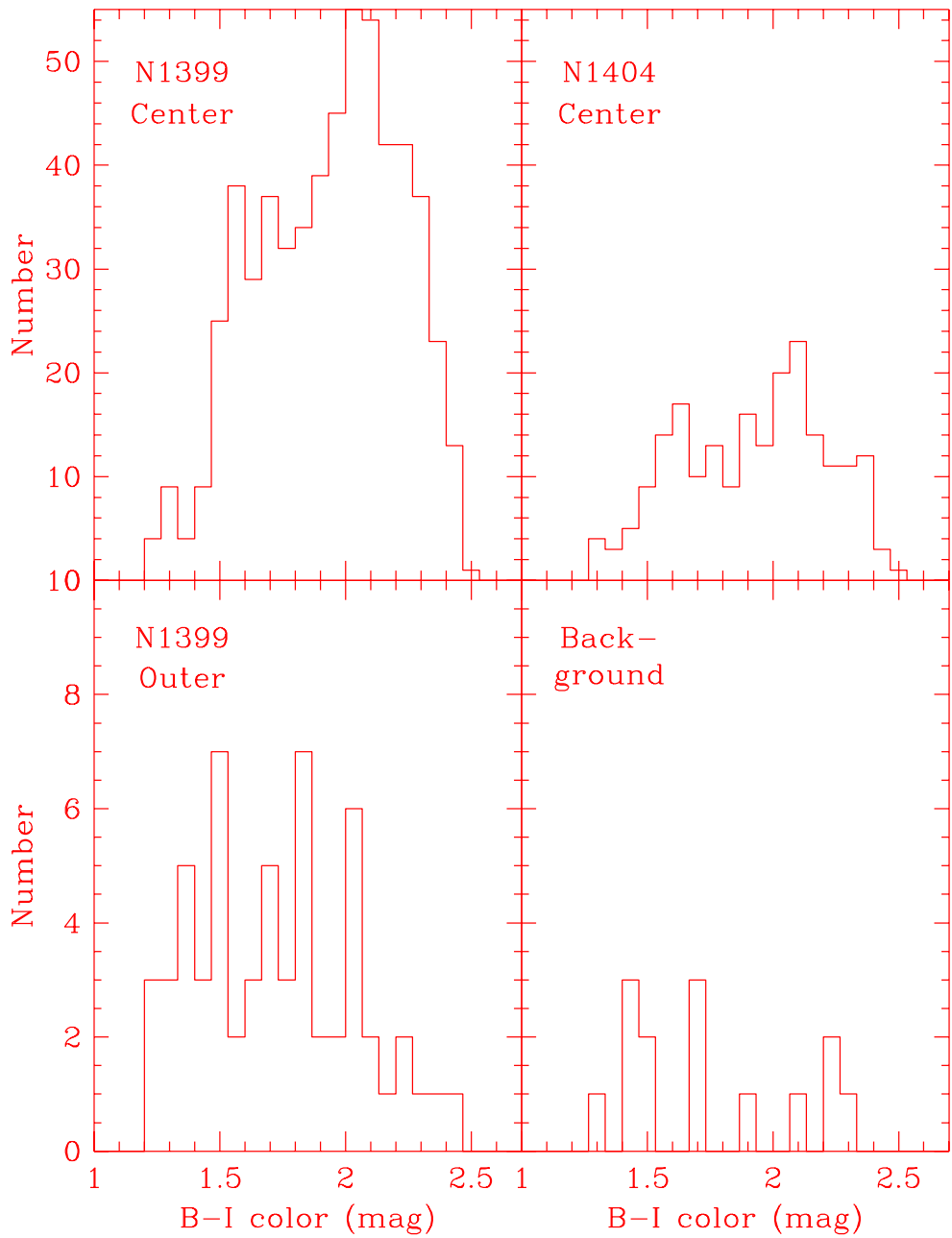,width=400pt}}
\caption{\label{fig6}
Histograms of globular cluster B--I colours for the four HST
pointings. 
In the central pointings, both galaxies are dominated by
red globular clusters with an extended blue tail.
}
\end{figure*}

\begin{figure*} 
\centerline{\psfig{figure=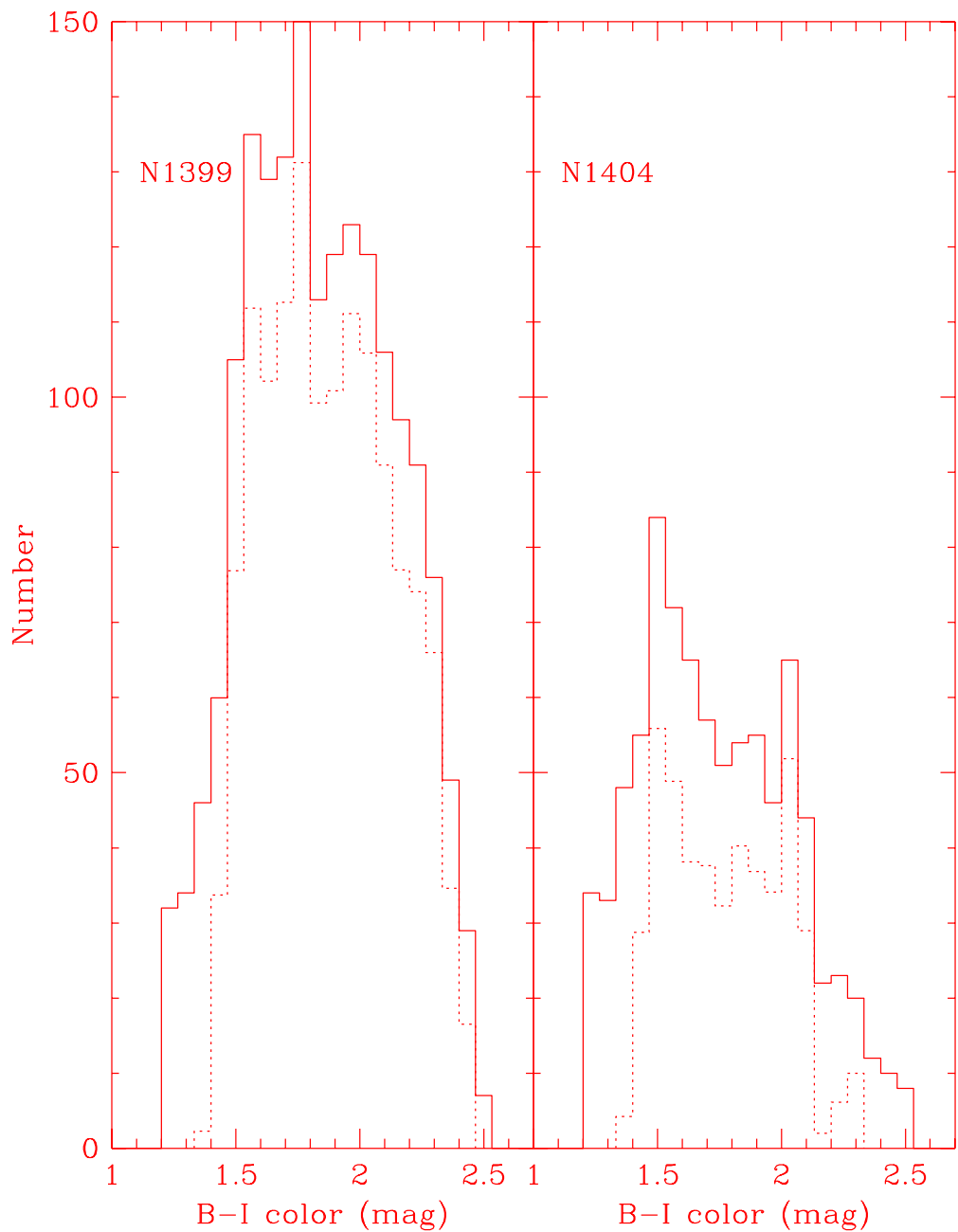,width=400pt}}
\caption{\label{fig7}
Histograms of globular cluster B--I colours for the two CTIO 
pointings. 
The dashed line shows the distribution after
correction for background contamination. 
In these wide area pointings, both galaxies are dominated by
blue globular clusters with an extended red tail.
}
\end{figure*}

\begin{figure*} 
\centerline{\psfig{figure=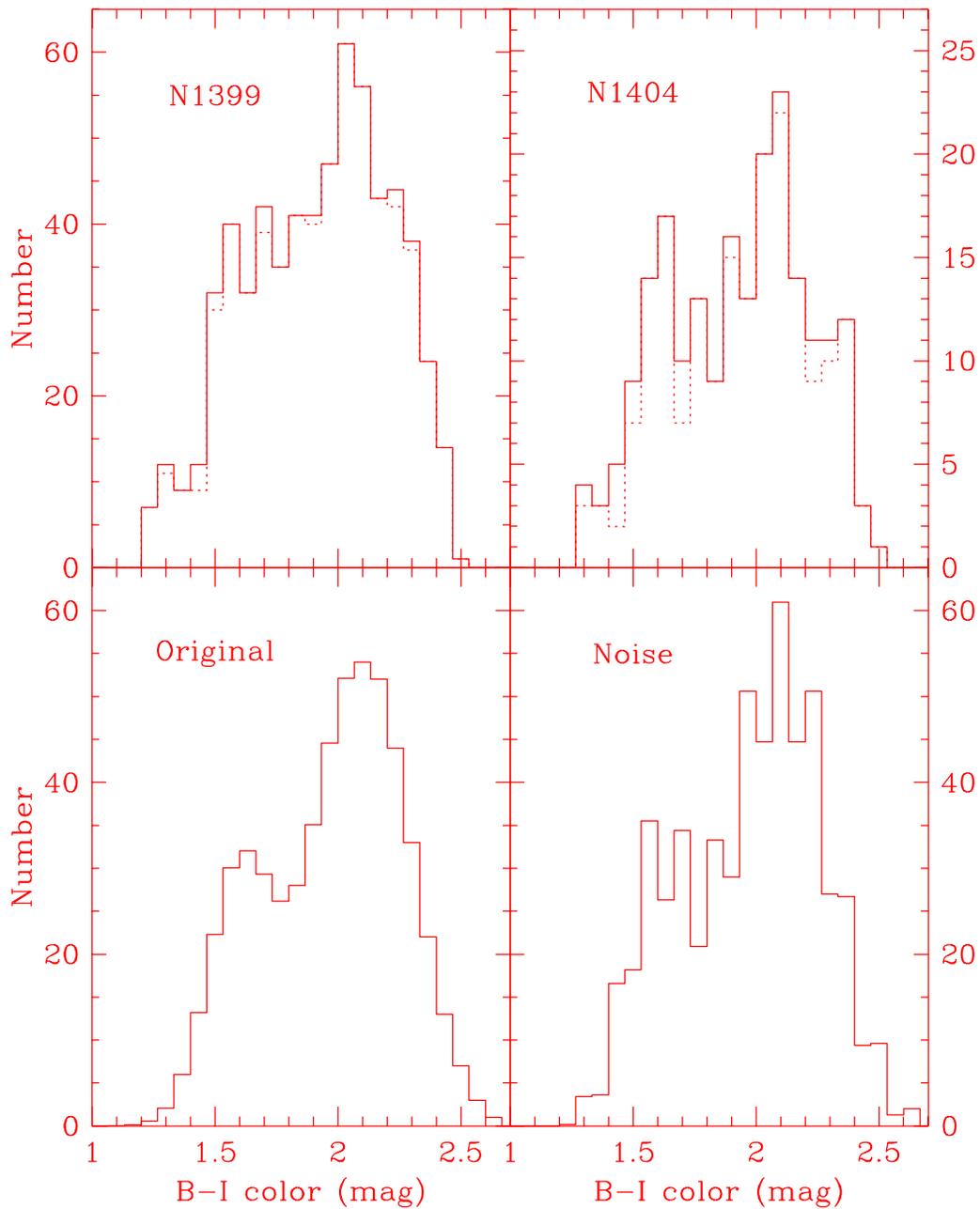,width=400pt}}
\caption{\label{fig8}
Histograms of the globular cluster B--I colours in NGC 1399 and NGC 1404
from HST data (upper panels). 
For NGC 1399 the central and outer pointings have been co--added. 
The dashed line shows the distribution after
correction for background contamination. The lower left panel shows the sum
of two Gaussians with peaks at B--I = 1.6 and 2.1. The lower right panel
shows the same two Gaussians with Poisson noise included. 
The metallicity distribution of 
both galaxies can be plausibly represented by two Gaussians.
} 
\end{figure*}

\begin{figure*} 
\centerline{\psfig{figure=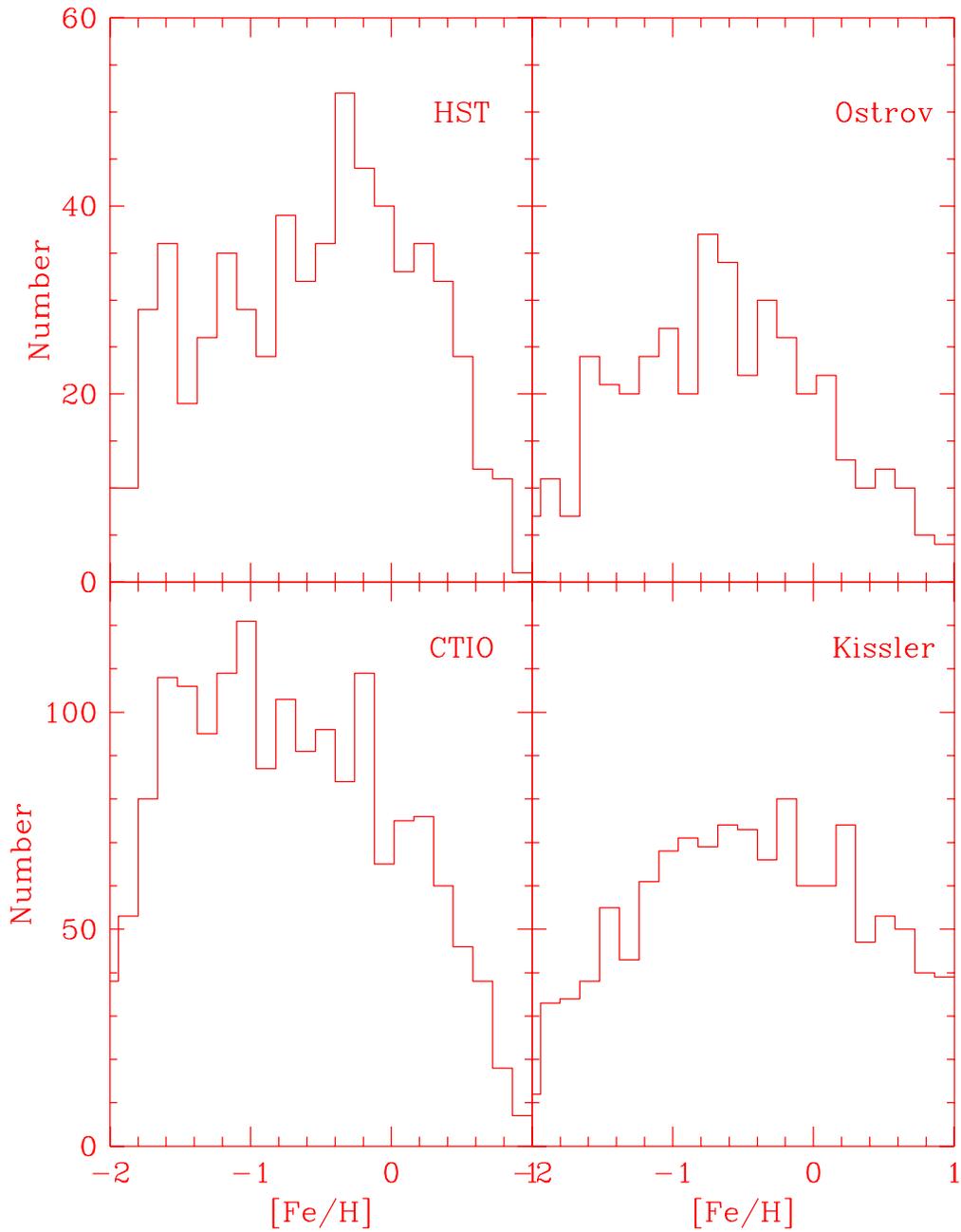,width=400pt}}
\caption{\label{fig9}
The metallicity distribution for globular clusters in NGC 1399 from four
different samples. Data from this paper are shown in the left hand side panels.
The upper right panel shows the data of Ostrov \etal (1993), and
the lower right panel shows the data of Kissler--Patig \etal (1997a).
None of the samples shown here 
have been background corrected (although for the
HST sample this is negligible). The Ostrov \etal data have the highest
metallicity sensitivity. 
The HST distribution is inconsistent with a
single Gaussian and is better described as multimodal. 
Common peaks (particularly at [Fe/H] $\sim$ --0.2)
can be seen in each data set.  
} 
\end{figure*}

\begin{figure*} 
\centerline{\psfig{figure=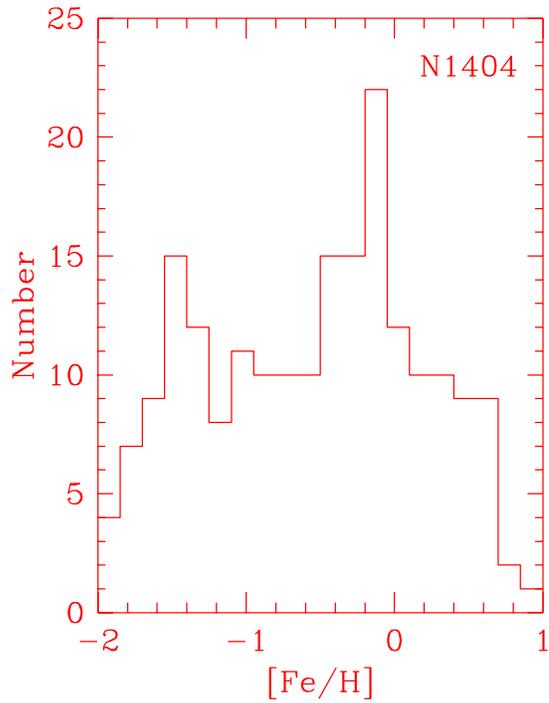,width=400pt}}
\caption{\label{fig10}
The metallicity distribution for globular clusters in NGC
1404 from HST data. A KMM test indicates that the distribution is
inconsistent with a single Gaussian and that a bimodal distribution with
mean metallicities of [Fe/H] $\sim$ --1.5 and --0.1 provides a better fit. 
}
\end{figure*}

\begin{figure*} 
\centerline{\psfig{figure=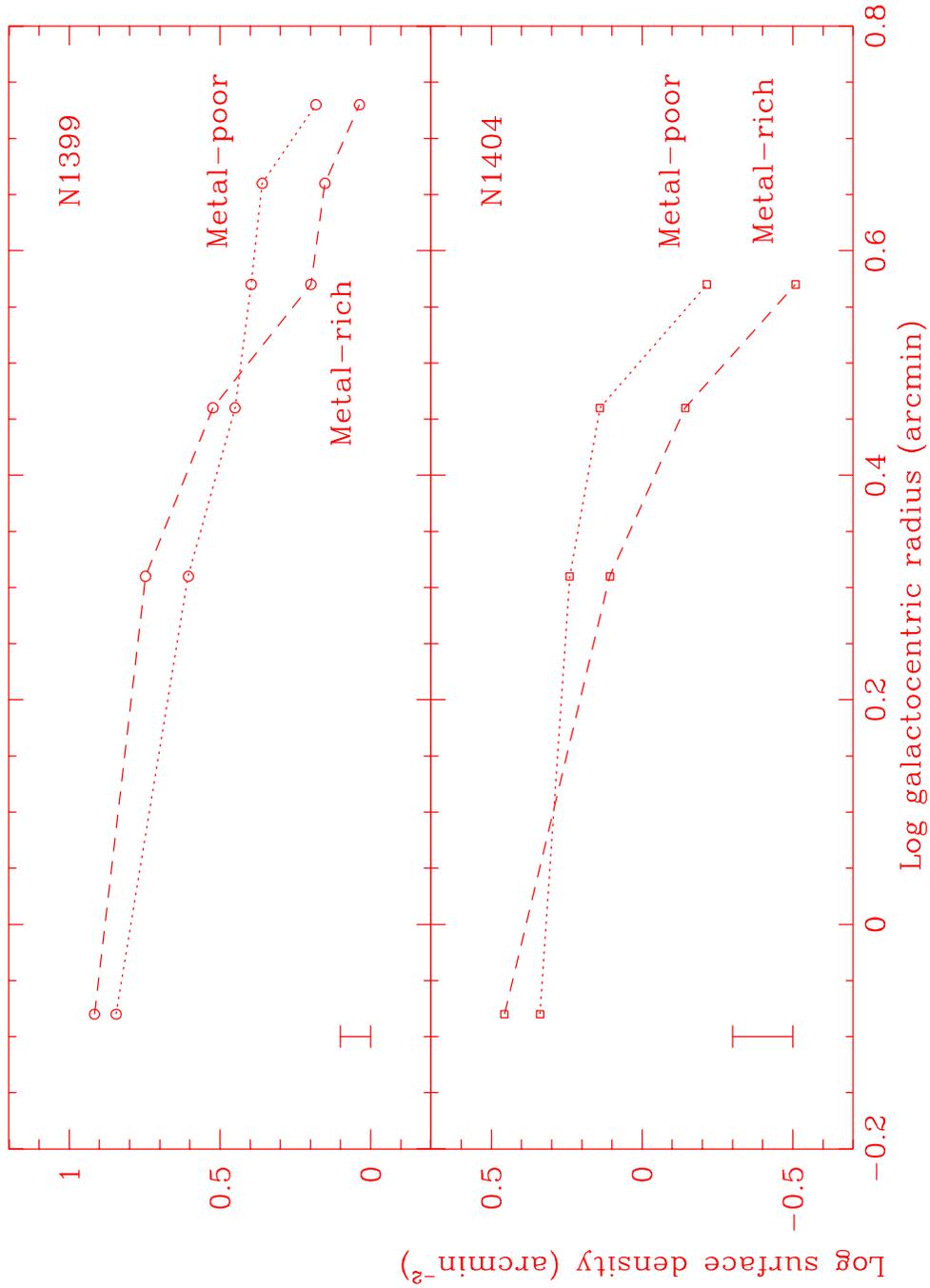,width=400pt}}
\caption{\label{fig11}
Surface density profiles for the metal--rich and metal--poor globular
cluster subpopulations in NGC 1399 (circles) and NGC 1404 (squares) from
CTIO data. No correction has been applied for background contamination or
missing globular clusters at the faint end of the luminosity function. 
Typical data error bars are shown on the left. In both galaxies the two
subpopulations decline with galactocentric radius indicating that they are
dominated by globular clusters. The metal--rich subpopulation is more
centrally concentrated. 
} 
\end{figure*}

\begin{figure*} 
\centerline{\psfig{figure=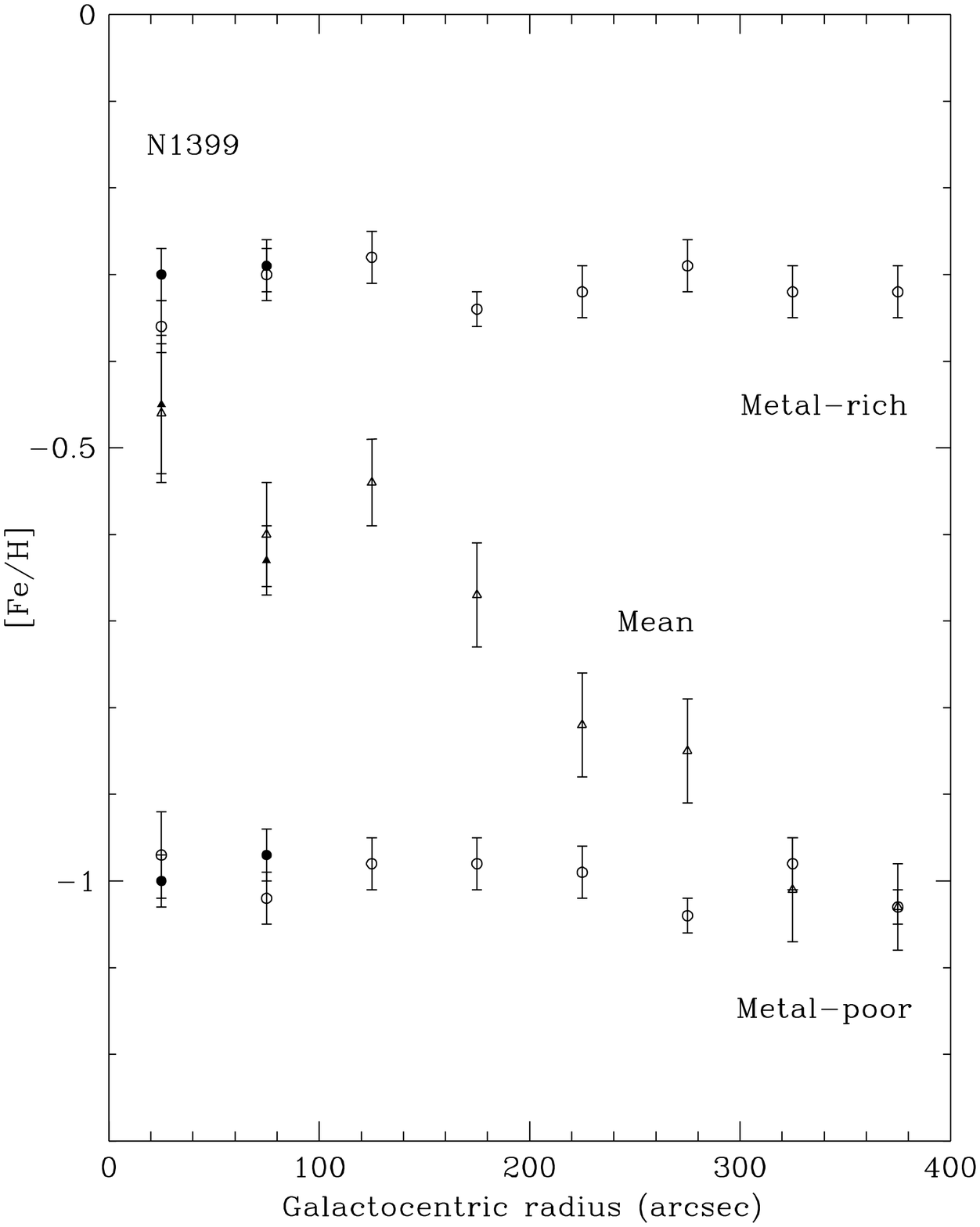,width=400pt}}
\caption{\label{fig12}
Radial variation of metallicity for the globular cluster
subpopulations in NGC 1399. The filled circles show the HST data and
the open circles the CTIO data. The metal--rich, 
metal--poor and mean for the whole globular cluster system are shown.
The mean metallicity gradient is consistent with the changing relative
proportions of the subpopulations with galactocentric radius.
}
\end{figure*}

\begin{figure*} 
\centerline{\psfig{figure=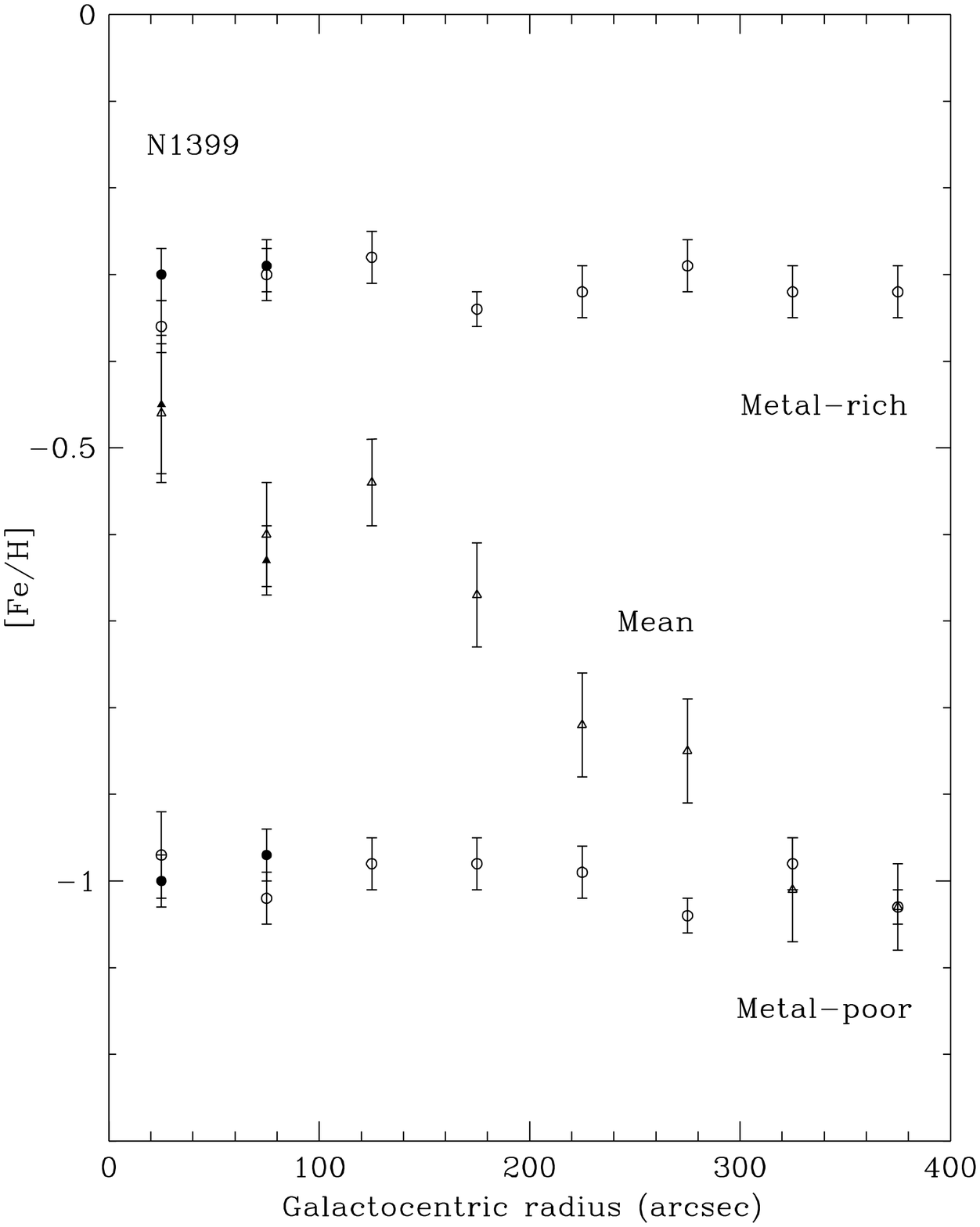,width=400pt}}
\caption{\label{fig13}
Radial variation of metallicity for the globular cluster
subpopulations in NGC 1404. The filled circles show the HST data and
the open circles the CTIO data. The metal--rich, intermediate,
metal--poor and mean for the whole globular cluster system are shown.
The mean metallicity gradient is consistent with the changing relative
proportions of the subpopulations with galactocentric radius.
}
\end{figure*}

\normalsize


\label{lastpage}

\end{document}